\documentclass[10pt,letterpaper]{article}
\usepackage[utf8]{inputenc}
\usepackage{amsmath}
\usepackage{amsfonts}
\usepackage{amssymb}
\usepackage{multirow}
\usepackage{bm}
\usepackage{xcolor}
\usepackage[english]{babel}
\usepackage{graphicx}
\usepackage{hyperref,cite}
\hypersetup{
    colorlinks=true,
    linkcolor=blue,
    filecolor=blue,  
    citecolor=blue,
    urlcolor=blue,
    }
\usepackage{lipsum}

\title{Generation of Standing Waves on a Real String}

\author{J. F. Pérez-Barragán\footnote{Contact email address: jfcoperezb@estudiantes.fisica.unam.mx}\\
\normalsize{Instituto de Física, Universidad Nacional Autónoma de México, 04510 Mexico City, Mexico}}

\date{}

\begin{document}

\maketitle

\abstract{\noindent We investigate the generation of standing waves in the model provided by the inhomogeneous telegraph equation under different forcing conditions. We show that sustained standing waves arise only for a specific forcing that is spatially distributed, continuous, and resonant.

\vspace{0.25cm}
\noindent \textbf{Keywords:} Telegraph equation, fourth-order spatial derivative, standing waves, forced oscillations, resonance}

\section{Introduction}\label{S1}

Standing waves,
\begin{equation}
f\sin(\omega t)\sin(kx),\label{StandingW}
\end{equation}
where $f$, $k$, and $\omega=ck$ denote the amplitude, wave number, and angular frequency, respectively, with $c$ the speed of traveling waves, are fundamental in many areas of physics, as they arise naturally in systems where waves are confined to finite domains. Canonical examples of these waves include acoustic waves in wind instruments, electromagnetic waves in resonant cavities, and wavefunctions of stationary quantum states in an infinite potential well. The simplest example, however, is that of mechanical waves on a stretched string.

Despite their ubiquity, one aspect of standing waves that has yet to be addressed is their generation under real-world conditions. In nature, the propagation of waves cannot be fully described by the wave equation, since real systems always involve additional phenomena that disturb the wave. For instance, in a stretched string, the surrounding air damps the oscillations, while internal friction dissipates energy. Moreover, real strings are not perfectly flexible, as they exhibit bending stiffness. A more realistic model for the transverse displacement $u(x,t)$ of the string is thus provided by the equation
\begin{equation}
    u_{tt}+2\gamma{u}_{t}-c^{2} u_{xx}+\kappa{c}^{4}u_{xxxx}=0,\label{EcHomogenea}
\end{equation}
with $\gamma$ and $\kappa$ the effective damping and bending stiffness coefficients, respectively. This equation corresponds to the telegraph equation (without leakage) with an additional fourth-order spatial derivative term that accounts for bending stiffness \cite{Courant1962,Tikhonov1990,Evans2010}.

Since its original proposal by Heaviside in 1876 to describe the propagation of electrical signals in conducting wires \cite{OH1876}, the telegraph equation has found applications in a wide range of physical contexts. In particular, it governs heat transport at finite speeds in thermally conducting media \cite{CC1958,PV1958}, the propagation of electromagnetic waves in conducting or dissipative media \cite{Jackson1975}, and the dynamics of random motions with dichotomous stochastic velocities (telegraph processes) \cite{MK1974}. It has also been proposed as a generalization of the Schrödinger equation that incorporates finite propagation speeds \cite{PS1997}. Of special interest is the connection established by Gaveau et al. between this equation and the Dirac equation, which is obtained by precisely interpreting the latter in terms of a telegraph process \cite{BG1984}.

Similarly, fourth-order spatial derivative terms have arisen in various areas of physics beyond their original introduction in the theory of elasticity. Notable examples include their role in describing the smoothing of interfaces via surface diffusion \cite{WM1957}, spontaneous phase separation in binary fluids \cite{JC1958}, and the stabilization of optical solitons through higher-order dispersion \cite{VK1996}. More recently, proposals for quantum gravity, such as Hořava–Lifshitz gravity, have incorporated such terms to regulate short-distance fluctuations of spacetime \cite{PH2009}. Hence, although classical in origin, the telegraph equation and its extensions, such as Eq. (\ref{EcHomogenea}), remain relevant in modern physics, and their properties continue to be the subject of active investigation.

Now, standing waves cannot be sustained on their own in dissipative media. Nevertheless, it is conceivable that an appropriate forcing could maintain them by driving the system toward a steady state in which the energy dissipated by damping is continuously compensated by the external input. This work is devoted to proving the existence of such states in the model provided by the inhomogeneous telegraph equation. The article is organized as follows. Sect. \ref{S2} presents the construction of the general solution to this equation using Green’s functions. In Sect. \ref{S3}, it is shown that the generation of standing waves under realistic conditions requires a specific spatially extended, continuous, and resonant forcing. Sect. \ref{S4} is concerned with the analysis of solutions corresponding to more realistic forcing scenarios. The paper concludes with a brief discussion of the implications of the findings.

\section{The General Solution}\label{S2}

To study the real-world oscillations of a stretched string initially at rest and with ends fixed in position but offering no resistance to bending at $x=0$ and $x=L$, we consider the boundary-value problem that is defined by the inhomogeneous telegraph equation
\begin{equation}
    u_{tt}+2\gamma{u}_{t}-c^{2} u_{xx}+\kappa{c}^{4}u_{xxxx}=f(x,t),\label{EcInHomogenea}
\end{equation}
where $f(x,t)$ is the forcing term, and subject to the initial conditions $u(x,0)=0$ and $u_{t}(x,0)=0$, together with the boundary conditions $u(0,t)=u(L,t)=0$, that correspond to fixed ends, and $u_{xx}(0,t)=u_{xx}(L,t)=0$, that express the absence of bending resistance at the boundaries.

The Green's function $G(x,x',t,t')$ associated with Eq. (\ref{EcInHomogenea}) satisfies
\begin{equation}
    G_{tt}+2\gamma{G}_{t}-c^{2} G_{xx}+\kappa{c}^{4}G_{xxxx}=\delta(x-x')\delta(t-t'),\label{Greenfunction}
\end{equation}
together with the same initial and boundary conditions as $u(x,t)$. Then, by considering its Fourier transform with respect to $t$,
\begin{equation}
    \tilde{G}(x,x',\omega)=\int_{-\infty}^{\infty} G(x,x',t,t') e^{i\omega{t}}dt,
\end{equation}
we obtain
\begin{equation}
    \kappa{c}^{4}\tilde{G}_{xxxx}-c^{2}\tilde{G}_{xx}-c^{2}k^{2}\tilde{G}=e^{i\omega{t}'}\delta(x-x'),\label{EcGreenHelmholtz}
\end{equation}
with $c^{2}k^{2}=\omega^{2}-2i\gamma\omega$. Next, given that the eigenfunctions of the homogeneous Helmholtz equation with homogeneous Dirichlet boundary conditions at $x=0$ and $x=L$,
\begin{equation}
    \phi_{n}(x)=\sin(k_{n}x),
\end{equation}
where $k_{n}=n\pi/L$ and ${n}=1,2,3\dots$, satisfy  boundary conditions identical to those of our problem and form a complete orthogonal basis on $[0,L]$, we expand $\tilde{G}(x,x',\omega)$ as
\begin{equation}
    \tilde{G}(x,x',\omega)=\sum_{n=1}^{\infty}a_{n}\phi_{n}(x).
\end{equation}
Substituting this expansion into Eq. (\ref{EcGreenHelmholtz}) yields
\begin{equation}
    \sum_{n=1}^{\infty}a_{n}(\kappa{c}^{4}k_{n}^{4}+c^{2}k_{n}^{2}-c^{2}k^{2})\phi_{n}(x)= e^{i\omega{t}'}\delta(x-x'),
\end{equation}
and the coefficients $a_{n}$ directly follow from the orthogonality of the eigenfunctions.


The Green's function associated with Eq. (\ref{EcInHomogenea}) is thus given by the inverse Fourier transform of $\tilde{G}(x,x',\omega)$:
\begin{equation}
    G(x,x',t,t')=-\frac{1}{\pi{L}}\sum_{n=1}^{\infty}\phi_{n}(x)\phi_{n}(x')\int_{-\infty}^{\infty}\frac{e^{-i\omega(t-t')}d\omega}{\omega^{2}-2i\gamma\omega-\omega_{n}^{2}-\kappa\omega_{n}^{4}},
\end{equation}
with $\omega_{n}=ck_{n}$, which leads to
\begin{equation}
    G(x,x',t,t')=\frac{2}{L}\Theta(t-t')e^{-\gamma(t-t')}\sum_{n=1}^{\infty}\frac{1}{\omega_{\mathrm{eff}}(n)}\sin(\omega_{\mathrm{eff}}(n)(t-t'))\phi_{n}(x)\phi_{n}(x'),
\end{equation}
where $\Theta(t)$ denotes the Heaviside step function and $\omega_{\mathrm{eff}}(n)=\sqrt{\omega_{n}^{2}+\kappa\omega_{n}^{4}-\gamma^{2}}$. Consequently, by expanding the forcing term as
\begin{equation}
    f(x,t)=\sum_{n=1}^{\infty}f_{n}(t)\phi_{n}(x),
\end{equation}
we find that the solution of Eq. (\ref{EcInHomogenea}) is given by
\begin{equation}
    u(x,t)=u_{\mathrm{hom}}(x,t)+\sum_{n=1}^{\infty}\frac{1}{\omega_{\mathrm{eff}}(n)}\phi_{n}(x)\int_{0}^{\infty}e^{-\gamma{t}'}\sin(\omega_{\mathrm{eff}}(n)t')f_{n}(t-t')dt',\label{SolGral}
\end{equation}
with
\begin{equation}
    u_{\mathrm{hom}}(x,t)=e^{-\gamma{t}}\sum_{n=1}^{\infty}(A_{n}e^{i\omega_{\mathrm{eff}}(n)t}+B_{n}e^{-i\omega_{\mathrm{eff}}(n)t})\phi_{n}(x)
\end{equation}
the homogeneous part of the solution. Finally, the initial conditions imply
\begin{equation}
    A_{n}=-\frac{1}{2\omega_{\mathrm{eff}}^{2}(n)}\int_{0}^{\infty}e^{-\gamma{t}'}\sin(\omega_{\mathrm{eff}}(n)t')([\omega_{\mathrm{eff}}(n)-i\gamma]f_{n}(-t')-i\dot{f}_{n}(-t'))dt',
\end{equation}
\begin{equation}
    B_{n}=-
    \frac{1}{2\omega_{\mathrm{eff}}^{2}(n)}\int_{0}^{\infty}e^{-\gamma{t}'}\sin(\omega_{\mathrm{eff}}(n)t')([\omega_{\mathrm{eff}}(n)+i\gamma]f_{n}(-t')+i\dot{f}_{n}(-t'))dt',
\end{equation}
which totally determines the solution.

A key consequence of incorporating both the fourth-order spatial derivative and damping terms in Eq. (\ref{EcInHomogenea}) is the modification of the dispersion relation:
\begin{equation}
    \omega_{\mathrm{eff}}(n)=n\omega_{1}\sqrt{1+\alpha{n}^{2}+\beta{n}^{-2}},\label{DispersionRelation}
\end{equation}
with $\alpha=\kappa\omega_{1}^{2}$ and $\beta=(\gamma/\omega_{1})^{2}$. This dispersion relation implies a shift in the frequency spectrum, since higher harmonics are no longer integer multiples of the fundamental frequency. This effect, known as inharmonicity, is commonly observed in musical instruments, where tuning procedures must account for such deviations \cite{Berg1982}. From Eq. (\ref{DispersionRelation}), we observe that the harmonic behavior is recovered in the simultaneous limits $\kappa \ll \omega_{1}^{-2}$ and $\gamma \ll \omega_{1}$, corresponding to the low-stiffness and weak-damping regimes, respectively. Notably, the inharmonicity induced by damping diminishes with increasing $n$, so that its effect becomes progressively less significant at higher frequencies. By contrast, the stiffness contribution increases quadratically with $n$, leading to increasingly pronounced deviations from harmonicity. As a result, even for small bending stiffness, the harmonic spectrum of an ideal string is only approximately reproduced by the lowest normal modes of a real string.

\section{Generation of Standing Waves}\label{S3}

We now address the problem of determining the form that $f_{n}(t)$ must take in order for the solution of Eq. (\ref{SolGral}) to reduce to the standing-wave form given in Eq. (\ref{StandingW}). Motivated by the physics of a forced damped harmonic oscillator, we consider the resonant forcing that results from taking
\begin{equation}
    f_{n}(t)=C_{n}\sin(\omega_{n}t)+D_{n}\cos(\omega_{n}t).
\end{equation}
Consequently, substituting into the general solution of Eq. (\ref{SolGral}) leads to
\begin{equation}
    u(x,t)=u_{\mathrm{hom}}(x,t)+\sum_{n=1}^{\infty}\frac{(\kappa\omega_{n}^{3}C_{n}+2\gamma{D}_{n})\sin(\omega_{n}t)+(\kappa\omega_{n}^{3}D_{n}-2\gamma{C}_{n})\cos(\omega_{n}t)}{\omega_{n}(\kappa^{2}\omega_{n}^{6}+4\gamma^{2})}\phi_{n}(x).
\end{equation}
By choosing $C_{n}=\kappa\omega_{n}^{4}f_{n}$ and ${D}_{n}=2\gamma\omega_{n}{f}_{n}$, we thus obtain that the forcing
\begin{equation}
    f(x,t)=\sum_{n=1}^{\infty}f_{n}\omega_{n}\sqrt{\kappa^{2}\omega_{n}^{6}+4\gamma^{2}}\sin(\omega_{n}t+\varphi_{n})\phi_{n}(x),\label{ForcingSW}
\end{equation}
with $\varphi_{n}=\arctan(2\gamma/\kappa\omega_{n}^{3})$, implies
\begin{equation}
    u(x,t)=\sum_{n=1}^{\infty}f_{n}\left(\sin(\omega_{n}t)-\frac{\omega_{n}}{\omega_{\mathrm{eff}}(n)}e^{-\gamma{t}}\sin(\omega_{\mathrm{eff}}(n)t)\right)\phi_{n}(x).\label{SupSW}
\end{equation}
This solution shows that by selecting a single frequency via $f_{n}=f\delta_{mn}$, with $\delta_{mn}$ the Kronecker delta, the forcing term resulting from Eq. (\ref{ForcingSW}) produces a response that converges, in the time-asymptotic limit, to the standing wave associated with the $m$-th harmonic of an ideal string. Therefore, we conclude that for generating standing waves on a real string, a continuous forcing tuned to the frequency of the desired normal mode and matching its spatial profile is required.

We can also analyze the evolution of the string's oscillations toward the steady-state regime from an energetic perspective. The mechanical energy of an ideal string is given by
\begin{equation}
    E(t)=\frac{\mu}{2}\int_{0}^{L}(u_{t}^{2}+c^{2}u_{x}^{2})dx,\label{Energy}
\end{equation}
where $\mu$ is the linear mass density \cite{Tikhonov1990}. Consequently, for the $n$-th harmonic contribution in Eq. (\ref{SupSW}), we have
\begin{equation}
    E(t)=\frac{\mu{L}}{4}f_{n}^{2}\omega_{n}^{2}(1+e^{-\gamma{t}}S_{1}(t)+e^{-2\gamma{t}}S_{2}(t)),\label{EnergySW}
\end{equation}
with
\begin{equation}
    S_{1}(t)=\frac{2}{\omega_{\mathrm{eff}}(n)}([\gamma\cos(\omega_{n}t)-\omega_{n}\sin(\omega_{n}t)]\sin(\omega_{\mathrm{eff}}(n)t)-\omega_{\mathrm{eff}}(n)\cos(\omega_{n}t)\cos(\omega_{\mathrm{eff}}(n)t)),\label{FunS1}
\end{equation}
\begin{equation}
    S_{2}(t)=\frac{1}{\omega_{\mathrm{eff}}^{2}(n)}(\omega_{n}^{2}\sin^{2}(\omega_{n}t)+[\omega_{\mathrm{eff}}(n)\cos(\omega_{\mathrm{eff}}(n)t)-\gamma\sin(\omega_{\mathrm{eff}}(n)t)]^{2}).\label{FunS2}
\end{equation}
Hence, $E(t)$ oscillates while approaching the constant value $\mu{L}f_{n}^{2}\omega_{n}^{2}/4$, which corresponds to the energy of a standing wave of amplitude $f_{n}$ and frequency $\omega_{n}$. On this basis, we conclude that when the energy of the string takes a constant value, the power supplied by the forcing balances the power dissipated by damping, and the system thereby reaches a steady state.

\section{More Realistic Forcing Scenarios}\label{S4}

Although effectively generating standing waves, a continuous forcing acting over the entire length of the string is difficult to realize in practice. We therefore examine two more realistic scenarios.

\subsection{Spatially Distributed, Impulsive Forcing}

We first consider a spatially distributed forcing that is applied on the string at a single instant, namely at $t_{0}$, which can be modeled by taking
\begin{equation}
    f_{n}(t)=C_{n}\delta(t-t_{0}).\label{ForcingDeltaT}
\end{equation}
Accordingly, Eq. (\ref{SolGral}) leads to the solution
\begin{equation}
    u(x,t)=\Theta(t-t_{0})e^{-\gamma(t-t_{0})}\sum_{n=1}^{\infty}\frac{C_{n}}{\omega_{\mathrm{eff}}(n)}\sin(\omega_{\mathrm{eff}}(n)(t-t_{0}))\phi_{n}(x).\label{DecaySW}
\end{equation}
Thus, by redefining $C_{n}$ as $f_{n}\omega_{\mathrm{eff}}(n)$, we have that the solution in Eq. (\ref{DecaySW}) can be written in a form resembling the time-asymptotic limit of Eq. (\ref{SupSW}), with the important differences that its components oscillate at frequency $\omega_{\mathrm{eff}}(n)$, rather than $\omega_{n}$, and that its amplitude decays exponentially with time. However, in the low-stiffness and weak-damping regimes, we have, by selecting a single frequency, that the forcing term in Eq. (\ref{ForcingDeltaT}) generates a response that corresponds to the standing wave of one of the lowest normal modes of an ideal string, but whose amplitude continuously decreases.

The energy associated with the $n$-th harmonic contribution in Eq. (\ref{DecaySW}) is given by
\begin{equation}
    E(t)=\frac{\mu{L}}{4}{f}_{n}^{2}\Theta(t-t_{0})e^{-2\gamma(t-t_{0})}S_{3}(t-t_{0}),
\end{equation}
with
\begin{equation}
    S_{3}(t)=\omega_{n}^{2}\sin^{2}(\omega_{\mathrm{eff}}(n)t)+(\omega_{\mathrm{eff}}(n)\cos(\omega_{\mathrm{eff}}(n)t)+\gamma\sin(\omega_{\mathrm{eff}}(n)t))^{2}.
\end{equation}
As expected, the energy also decays exponentially for the oscillations generated by the forcing in Eq. (\ref{ForcingDeltaT}). It is revealing, nonetheless, that in the low-stiffness and weak-damping regimes,
\begin{equation}
    E(t)\approx\frac{\mu{L}}{4}{f}_{n}^{2}\omega_{n}^{2}\Theta(t-t_{0})e^{-2\gamma(t-t_{0})},
\end{equation}
which can be interpreted as the mechanical energy of a standing wave of amplitude $f_{n}e^{-\gamma(t-t_{0})}$ and frequency $\omega_{n}$.

\subsection{Spatially Localized, Continuous Forcing}

In contrast, we now consider a forcing acting continuously at a fixed point of the string, namely at $x_{0}$. Specifically, we adopt the resonant forcing term
\begin{equation}
    f(x,t)=(C_{n}\sin(\omega_{n}{t})+D_{n}\cos(\omega_{n}{t}))\delta(x-x_{0}).
\end{equation}
Then, the corresponding solution is
\begin{equation}
    u(x,t)=u_{\mathrm{hom}}(x,t)+\frac{2}{L}\sum_{m=1}^{\infty}\phi_{m}(x_{0})([a_{mn}C_{n}+b_{mn}D_{n}]\sin(\omega_{n}{t})+[a_{mn}D_{n}-b_{mn}C_{n}]\cos(\omega_{n}{t}))\phi_{m}(x),\label{SolSLCForcing}
\end{equation}
where the coefficients $a_{mn}$ and $b_{mn}$ are given by
\begin{equation}
    a_{mn}=\frac{\omega_{m}^{2}-\omega_{n}^{2}+\kappa\omega_{m}^{4}}{(\omega_{m}^{2}-\omega_{n}^{2}+\kappa\omega_{m}^{4})^{2}+4\gamma^{2}\omega_{n}^{2}},
\end{equation}    
\begin{equation}    
    {b}_{mn}=\frac{2\gamma\omega_{n}}{(\omega_{m}^{2}-\omega_{n}^{2}+\kappa\omega_{m}^{4})^{2}+4\gamma^{2}\omega_{n}^{2}},
\end{equation}
respectively. 

For $m=n$, we have that the inhomogeneous part of the solution in Eq. (\ref{SolSLCForcing}) reduces to
\begin{equation}
    u_{\mathrm{inh}}^{m=n}(x,t)=\frac{2\phi_{n}(x_{0})([\kappa\omega_{n}^{3}C_{n}+2\gamma{D}_{n}]\sin(\omega_{n}{t})+[\kappa\omega_{n}^{3}D_{n}-2\gamma{C}_{n}]\cos(\omega_{n}{t}))}{\omega_{n}L(\kappa^{2}\omega_{n}^{6}+4\gamma^{2})}\phi_{n}(x).
\end{equation}
Hence, by choosing $C_{n}=\kappa\omega_{n}^{4}Lf/2\phi_{n}(x_{0})$ and $D_{n}=\gamma\omega_{n}Lf/\phi_{n}(x_{0})$, we have that
\begin{equation}
    f(x,t)=\frac{f\omega_{n}L}{2\phi_{n}(x_{0})}\sqrt{\kappa^{2}\omega_{n}^{6}+4\gamma^{2}}\sin(\omega_{n}t+\varphi_{n})\delta(x-x_{0}),\label{ForcingSWdelta}
    \end{equation} 
and that $u_{\mathrm{inh}}^{m=n}(x,t)$ thereby takes the form of a standing wave with frequency $\omega_{n}$. Consequently, Eq. (\ref{SolSLCForcing}) takes the form
\begin{equation}
    u(x,t)=f\sum_{m=1}^{\infty}\left(c_{mn}\sin(\omega_{n}{t})-\frac{\omega_{n}c_{mn}+\gamma{d}_{mn}}{\omega_{\mathrm{eff}}(m)}e^{-\gamma{t}}\sin(\omega_{\mathrm{eff}}(m)t)+d_{mn}[\cos(\omega_{n}{t})-e^{-\gamma{t}}\cos(\omega_{\mathrm{eff}}(m)t)]\right)\phi_{m}(x),\label{SolFinal4.2}
\end{equation}
with
\begin{equation}
    c_{mn}=\frac{\kappa\omega_{n}^{4}(\omega_{m}^{2}-\omega_{n}^{2}+\kappa\omega_{m}^{4})+4\gamma^{2}\omega_{n}^{2}}{(\omega_{m}^{2}-\omega_{n}^{2}+\kappa\omega_{m}^{4})^{2}+4\gamma^{2}\omega_{n}^{2}}\left[\frac{\phi_{m}(x_{0})}{\phi_{n}(x_{0})}\right],\label{CoefC}
\end{equation}
\begin{equation}
    d_{mn}=\frac{2\gamma\omega_{n}(\omega_{m}^{2}-\omega_{n}^{2}+\kappa(\omega_{m}^{4}-\omega_{n}^{4}))}{(\omega_{m}^{2}-\omega_{n}^{2}+\kappa\omega_{m}^{4})^{2}+4\gamma^{2}\omega_{n}^{2}}\left[\frac{\phi_{m}(x_{0})}{\phi_{n}(x_{0})}\right].\label{CoefD}
\end{equation}

From Eq. (\ref{SolFinal4.2}), we conclude that a spatially localized, continuous forcing cannot generate pure standing waves, since it irremediably excites multiple normal modes of the string. However, the forms of Eqs. (\ref{CoefC}) and (\ref{CoefD}) indicate that in the low-stiffness and weak-damping regimes,
\begin{equation}
    c_{mn}\sim\left(\frac{\gamma}{\omega_{1}}\right)^{2},\quad{d}_{mn}\sim\frac{\gamma}{\omega_{1}},
\end{equation}
so that the components of the solution with $m\neq{n}$ can be regarded as small perturbations. Therefore, the forcing term in Eq. (\ref{ForcingSWdelta}) acts as a source of approximate standing waves under these conditions.

Finally, the energy associated with Eq. (\ref{SolFinal4.2}) is given by
\begin{equation}
    E(t)=\frac{\mu{L}}{4}f^{2}\omega_{n}^{2}(1+e^{-\gamma{t}}S_{1}(t)+e^{-2\gamma{t}}S_{2}(t))+\frac{\mu{L}}{2}f^{2}\sum_{m\neq{n}}(\dot{S}_{4}^{2}(t)+\omega_{m}^{2}{S}_{4}^{2}(t)),\label{EnergySWSec4}
\end{equation}
where $S_{1}(t)$ and $S_{2}(t)$ are defined in Eqs. (\ref{FunS1}) and (\ref{FunS2}), respectively, whereas $S_{4}(t)$  corresponds to the time-dependent function in parentheses in Eq. (\ref{SolFinal4.2}). In contrast to the previous cases, the energy remains oscillatory. In the time-asymptotic limit, nevertheless, it oscillates around the mean value
\begin{equation}
    \langle{E(t)}\rangle=\frac{\mu{L}}{4}f^{2}\omega_{n}^{2}\left(1+\sum_{m\neq{n}}\frac{\omega_{m}^{2}+\omega_{n}^{2}}{\omega_{n}^{2}}(c_{mn}^{2}+d_{mn}^{2})\right),
\end{equation}
which approximates the energy of a standing wave of amplitude $f$ and frequency $\omega_{n}$ in the low-stiffness and weak-damping regimes.


\section{Conclusions}

We have shown that the generation of standing waves under real-world conditions requires a continuous resonant forcing with the spatial profile of the desired normal mode. Under these conditions, the oscillations on a real string are able to reach a steady state in which the energy dissipated by damping is exactly balanced by the external forcing. In contrast, the analysis of a spatially distributed, impulsive forcing allowed us not only to reproduce Mersenne’s laws for the fundamental harmonic of a stretched string in the low-stiffness and weak-damping regimes, but also to explain the well-established observation that the excitation produced by plucking or striking a string of a musical instrument produces not a pure tone, but a time-decaying sound consisting of a superposition of harmonics \cite{Berg1982}.

Furthermore, the study of a spatially localized, continuous forcing showed that the textbook example of generating standing waves by driving one end of a string is more subtle than often assumed, as this process inevitably excites multiple normal modes, although an approximate standing wave may still be observed (see, for example, Halliday et al. \cite{Halliday2013}). Finally, the fact that the amplitudes of the three obtained forcing terms depend on the frequency of the standing wave is noteworthy, as it indicates that the excitation of higher-frequency modes in a real string requires stronger forcing than that of lower-frequency modes. This result is consistent with the proportionality between the mechanical energy of a standing wave and the square of its frequency. With these findings, we aim to contribute to a deeper understanding of the phenomena described by the telegraph equation and its extensions.
\vspace{0.5cm}

\noindent \textbf{Acknowledgments:} The author is grateful to Professor A. M. Cetto for bringing this problem to his attention and for her insightful comments.


\vspace{0.5mm}

\begin{thebibliography}{99}

\bibitem{Courant1962}
Courant, R., Hilbert, D.: Methods of Mathematical Physics. Volume II: Partial Differential Equations. John Wiley \& Sons, New York (1962)

\bibitem{Tikhonov1990}
Tikhonov, A. N., Samarskii, A. A.: Equations of Mathematical Physics. Dover Publications, Mineola, NY (1990)

\bibitem{Evans2010}
Evans, L. C.: Partial Differential Equations, 2nd ed. American Mathematical Society, Providence, RI (2010)

\bibitem{OH1876}
Heaviside, O.: XIX. On the Extra Current. \textit{Philos. Mag. S5} \textbf{2}, 135 (1876)

\bibitem{CC1958}
Cattaneo, C.: Sur une forme de l’equation de la chaleur eliminant le paradoxe d’une propagation instantanee. \textit{C. R. Acad. Sci.} \textbf{247}, 431 (1958)

\bibitem{PV1958}Vernotte, P.: La véritable équation de la chaleur. \textit{C. R. Acad. Sci.} \textbf{247}, 2103 (1958)

\bibitem{Jackson1975} 
Jackson, J. D.: Classical Electrodynamics, 2nd ed. John Wiley \& Sons, New York (1975)

\bibitem{MK1974}
Kac, M.: A Stochastic Model Related to the Telegrapher's Equation. \textit{Rocky Mountain J. Math.} \textbf{4}, 497 (1974)

\bibitem{PS1997}
Sancho, P.: The quantum telegraph equation. \textit{Nuovo Cim. B} \textbf{112}, 1437 (1997)

\bibitem{BG1984}
Gaveau, B., Jacobson, T., Kac, M., Schulman, L. S.: Relativistic Extension of the Analogy Between Quantum Mechanics and Brownian Motion. \textit{Phys. Rev. Lett.} \textbf{53}, 419 (1984)

\bibitem{WM1957}
Mullins, W. W.: Theory of Thermal Grooving. \textit{J. Appl. Phys.} \textbf{28}, 333 (1957)

\bibitem{JC1958}
Cahn, J. W., Hilliard, J. E.: Free Energy of a Nonuniform System. I. Interfacial Free
Energy. \textit{J. Chem. Phys.} \textbf{28}, 258 (1958)

\bibitem{VK1996}
Karpman, V. I.: Stabilization of soliton instabilities by higher-order dispersion: Fourth-order nonlinear Schrödinger-type equations. \textit{Phys. Rev. E} \textbf{53}, R1336 (1996)

\bibitem{PH2009}
Hořava, P.: Quantum gravity at a Lifshitz point. \textit{Phys. Rev. D} \textbf{79}, 084008 (2009)

\bibitem{Berg1982} 
Berg, R. E., Stork, D. G.: The Physics of Sound, 3rd ed. Pearson, Boston (2004)

\bibitem{Halliday2013} 
Halliday, D., Resnick, R., Walker, J.: Fundamentals of Physics, 10th ed. Wiley, Hoboken, NJ (2013)

\end{thebibliography}
\end{document}